\documentclass[11pt,prd,onecolumn,amsfonts,amssymb,amsmath,showkeys,nofootinbib,preprintnumbers]{revtex4-1}
\usepackage[english]{babel}
\usepackage[utf8]{inputenc}
\usepackage[colorinlistoftodos, color=green!40, prependcaption]{todonotes}
\usepackage{amsthm}
\usepackage{mathtools}
\usepackage{physics}
\usepackage{xcolor}
\usepackage{graphicx}
\usepackage{adjustbox}
\usepackage{placeins}
\usepackage[T1]{fontenc}
\usepackage{lipsum}
\usepackage{csquotes}
\usepackage[pdftex, pdftitle={Article}, pdfauthor={Author}]{hyperref} % For hyperlinks in the PDF
\usepackage{multirow}
\usepackage{array}
\usepackage{subcaption}

\newcommand{\sat}{\mathrm{sat}}
\newcommand{\sym}{\mathrm{sym}}
\newcommand{\qyc}{\mathrm{qyc}}

\begin{document}

\preprint{NP3M-P2300001}
\preprint{LA-UR-23-20878}

\title{Impact of O4 future detection on the determination of the dense matter equations of state}

\author{J.-F. Coupechoux$^{1}$}
\email{j-f.coupechoux@ip2i.in2p3.fr}
\author{R. Chierici$^1$}
\author{H. Hansen$^1$}
\author{J. Margueron$^1$}
\author{R. Somasundaram$^{1,2,3}$}
\author{V. Sordini$^1$}

\affiliation{$^1$Univ Lyon, Univ Claude Bernard Lyon 1, CNRS/IN2P3,\\
Institut de Physique des 2 Infinis de Lyon, UMR 5822, 69622 Villeurbanne, France}

\affiliation{$^2$Theoretical Division, Los Alamos National Laboratory, Los Alamos, New Mexico 87545, USA}

\affiliation{$^3$Department of Physics, Syracuse University, Syracuse, NY 13244, USA}

\date{\today} % Leave empty to omit a date

\begin{abstract}
In view of the next LIGO-Virgo-KAGRA Observing period O4 (to start in Spring 2023), we address the question of the ability of the interferometers network to discriminate among different neutron stars equation of states better than what was possible with the observation of the binary neutron stars merger GW170817. We show that the observation of an event similar to GW170817 during O4 would allow to resolve the dimensionless effective tidal deformability $\tilde{\Lambda}$ within an uncertainty 7 times better than the one obtained in O2. Thanks to the expected increase in sensitivities, we show that any GW170817-like single-event within a distance of 100 Mpc would imply significantly improved constraints of the neutron stars equations of state. We also illustrate the important impact of the noise in the analysis of the signal, showing how it can impact the effective tidal deformability probability density function for large signal-to-noise ratio.

\end{abstract}


\maketitle

\section{Introduction}

Neutron stars (NSs) are the densest compact objects known in the Universe. These stars have a radius of about ten to fourteen kilometers and masses observed between $1.174\pm 0.004 M_{\odot}$~\cite{Ozel:2016} and $2.14^{+0.10}_{-0.09} M_{\odot}$~\cite{Cromartie:2020} and they are the central residue of massive star collapse. The density in the core of a NS can reach up to about 8 times the nuclear saturation density (particle density: $n_\sat=0.155\pm0.005$~fm$^{-3}$~\cite{Margueron:2018a}, energy-density: $\rho_\sat\approx 2.7\times 10^{14}$~g~cm$^{-3}$). At such densities, the state of nuclear matter is still yet quite unknown, see Refs.~\cite{Weber:2004kj,Weber:2000xd} for some discussions, and it may undergo a phase transition between nuclear matter to some form of exotic matter. The composition of matter may therefore be nucleons, or quark-gluon plasma, or meson condensate, or hyperons, or H-dibaryon, etc. Astrophysical observations can bring improved knowledge on the composition of the core of NSs.

The structure of a hydrostatic spherical NS is determined by the Tolman-Oppenheimer-Volkoff (TOV) equation~\cite{Oppenheimer:1939ne} and the NS equation of state (EoS). The EoS of cold matter is assumed to be universal and hence to be the same for all NSs. From the solution of the TOV equation, one can determine the sequence of masses and radii allowed by a given EoS. The measurement of the mass and radii of pulsars is thus a primordial information to better constrain the possible NS EoS as done by the Neutron Star Interior Composition Explorer (NICER) project~\cite{Miller:2019cac,Riley:2021}.

The analyses of the GW170817 signal~\cite{LIGOScientific:2018cki} have shown the high potential of the detections of gravitational waves emitted by the merger of binary NS (BNS) systems in constraining the EoS in the core of a NS. 
In a binary system, each NS is under the action of the gravitational field of the companion star, $\mathcal{E}_{ij}$. As a consequence, the NS is tidally deformed and its quadrupole moment is equal to $Q_{ij}=-\Lambda \mathcal{E}_{ij}$, with $\Lambda$ the tidal deformability of the NS. The tidal deformability is a central parameter for the study of the EoS of cold nuclear matter at beta equilibrium and can be measured from GW signals. Provided the gravitational waveform (GW) is loud enough, it is possible to measure the tidal deformability with gravitational waves, and thus constrain the state of nuclear matter in NS.
This is possible because, during a coalescence of NSs, the effective tidal deformability of the BNS impacts the post-Newtonian (pN) waveform expansion at the fifth order.

The network of three Michelson interferometers of the LIGO-Virgo collaboration (LVC)~\cite{LIGOScientific:2014pky, VIRGO:2014yos} have already detected $90$ compact binary coalescences, most of which are black holes~\cite{LIGOScientific:2018mvr, LIGOScientific:2020ibl, LIGOScientific:2021djp}. The detection that better constrains the EoS is the event called GW170817 detected on August 17, 2017 at a distance of about 40 Mpc~\cite{LIGOScientific:2017vwq}. The gamma-ray burst GRB 170817A~\cite{LIGOScientific:2017zic} and the electromagnetic spectrum associated with this signal was also detected giving rise to the first multi-messenger study~\cite{LIGOScientific:2017ync}. The detection on Earth of the gravitational waves emitted by this exceptional event lead to the first measure of the tidal deformabilities from BNSs and thus put constraints on the EoS in a density region corresponding to the NS masses. For example, in paper~\cite{LIGOScientific:2018cki}, two scenarios for the EoS of matter were investigated. The first one is based on the so-called "insensitive EoS" relations~\cite{Yagi:2015pkc} and the connection between tidal deformability and the compactness of the NS~\cite{Yagi:2016bkt} to determine its radius. The second one directly assumes a spectral representation of the EoS, $p=p(\rho)$ linking the pressure $p$ and the energy density $\rho$, and depending on 4 parameters~\cite{Lindblom:2010bb}. This analysis not only constrains the NS EoS but also gives the probability density associated to these 4 parameters. Such an analysis has been repeated by several other groups, with different EoS representations and, for some of them, investigating the impact of electromagnetic counter parts, see for instance Refs.~\cite{De:2018,Capano:2019eae,Dietrich:2020,Pang:2021}.

In our study the estimation of the source properties from the gravitational wave signal is performed in a Bayesian framework. We are using Bilby~\cite{bilby_paper}, a parameter estimation framework for gravitational wave astronomy. In order to analyse a BNS GW signal, a relation between the tidal deformability and the mass should not be assumed. However, it is possible to assume a modeling of the EoS to calculate the tidal deformability, and use it to generate the corresponding GW. Instead of parameterizing a set of EoSs, if one just fixes the relation between the pressure and energy density, a value of the mass uniquely determines the value of the tidal deformability, and the parameters space explored by the Bayesian inference samplers has two dimensions less. In the  paper~\cite{LIGOScientific:2019eut}, 24 classical EoSs were considered to describe the state of matter of a NS. For each EoS, a Bayesian analysis has been performed. The selection of the best EoS and their ranking was done by the Bayes factor allowing to compare two competing models. In the present analysis we consider eight EoSs, six of them with an explicit transition to a quark core (first order or cross-over) in order to reanalyze GW170817. The aim of our new analysis is to explore the possibility of discriminating among EoSs with the O4 data, the next observing run of the LIGO-Virgo-KAGRA (LVK) collaboration, starting in Spring 2023. We study the impact of the expected noise reduction, compared to the O2 data when GW170817 was observed, as well as the impact of the source distance. %, and we evaluate the possibility of inferring the existence of a phase transition. 

The paper consists of two parts. The first part aims at reanalyzing GW170817 with the eight EoSs and rank them by the Bayes factor. We also present two different approximations to this end and discuss the benefit of each one of them. The second part of our study concerns  the possible discrimination between the EoS thanks to the observations in the O4 run. To do so, we use simulated signals injected on a noise with an expected power spectrum density corresponding to the O4 data taking. Given the expected sensitivity of KAGRA, we only consider LIGO and Virgo detectors for our simulation study. 

\section{Equations of state in the analysis of the GW170817 LVC data}

The main parameter carrying information about dense matter EoS in this context is the tidal deformability $\Lambda$.
As exceptional as the GW170817 event was, it has rejected only a small number of EoSs, most of which were already excluded by constraints coming from nuclear physics~\cite{Tews:2018iwm}. In the present study, we anticipate a much better capability of constraining the EoS during the next observational campaign O4, since the sensitivity will be substantially improved. The question that needs to be addressed is to understand to what extent the various predictions for the EoS can be discriminated, and what amount/quality of data will be needed to achieve this. To this end we have set-up a protocol where we assume a specific EoS while analysing the GW170817 data. We then compare the average uncertainties on the measurement of the tidal deformability parameters associated to such analyses to the uncertainty from an EoS-blind analysis, hereafter called \textsl{flat prior}. If the EoS-blind analysis allows us to exclude predictions based on specific EoSs, then the data will be able to distinguish amongst various EoSs. The ability of describing the data is assessed by the use of Bayes factors. 

The following section describes the set of EoSs that we have used in this study.

\subsection{Equations of state exploring different scenarios: nucleonic, first order phase transitions and quarkyonic cross-over}
\label{EOS}

\begin{figure}[t!]
\begin{subfigure}{0.49\textwidth}
    \includegraphics[width=\textwidth, trim = {40 0 50 5}, clip]{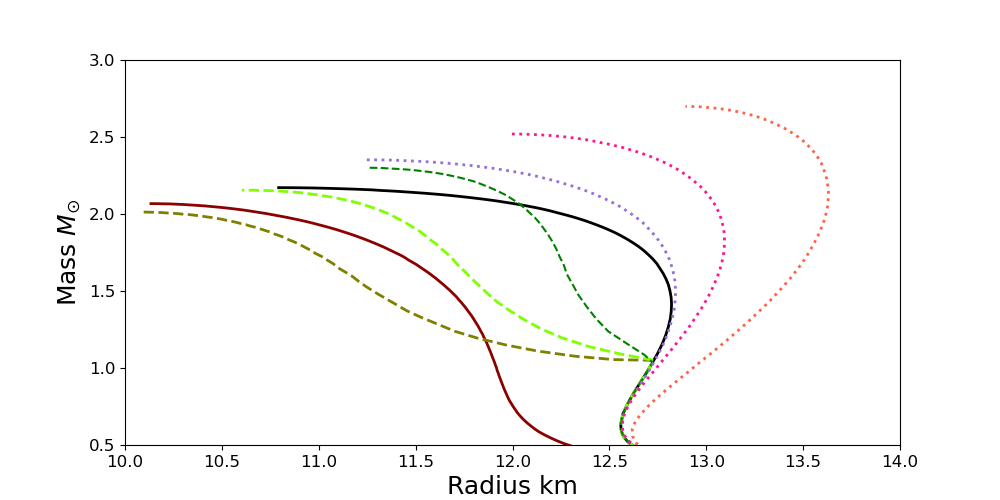}
    \caption{}
    \label{afig:rayon_lambda}
\end{subfigure}
\begin{subfigure}{0.49\textwidth}
    \includegraphics[width=\textwidth, trim = {40 0 50 5}, clip]{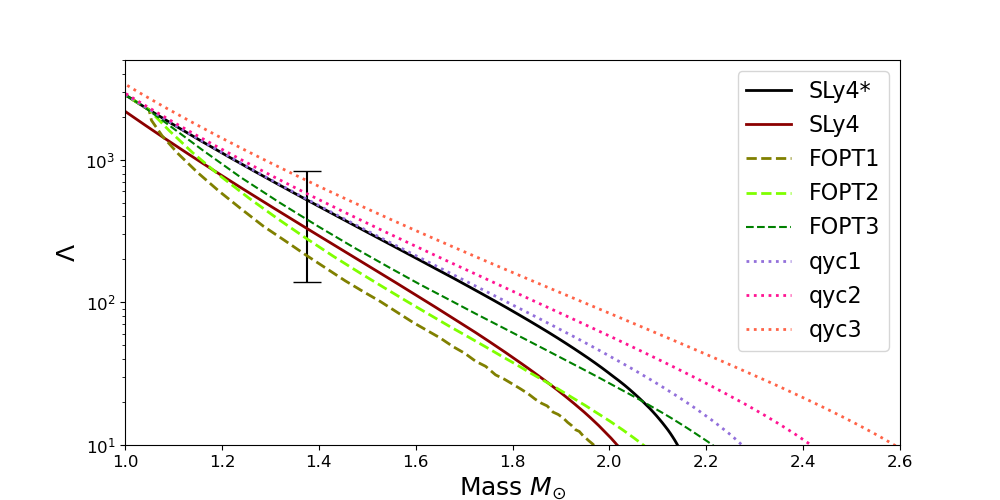}
    \caption{}
    \label{bfig:rayon_lambda}
\end{subfigure}
\caption{(a) The mass-radius curves for various EoSs considered in this analysis: SLy4 and SLy4$^*$ (solid lines) ; three EoSs with a FOPT (dashed lines) that presents a kink ; three qyc models (dotted lines). 
(b) The associated tidal deformabilities as a function of NS masses. For illustration, the vertical bar represents the 90$\%$ confidence interval of the tidal deformability $\tilde{\Lambda}$ extracted from the LVC analysis of GW170817~\cite{LIGOScientific:2018hze}.} 
\label{fig:rayon_lambda}
\end{figure}

The core of NSs can be composed of compressed neutrons and protons, but it can also be composed of deconfined quark matter. In the latter case, the star is an hybrid star where the inner and the outer core are separated by a phase transition. The article~\cite{Somasundaram:2021ljr} studies the impact of such a phase transition~\cite{Alford:2013aca} on the radius of the star and compares the predictions with the observations by NICER and the LVC (GW170817). This transition can either be first order, or quarks may be produced by a smooth cross-over process, such as the one suggested by the quarkyonic model (qyc) \cite{McLerran:2019} where the pressure does not present a kink as it is the case for a first order transition. In those quarkyonic stars one can distinguish an inner and outer core of different nature, but with no strict delimitation. 
For our analysis, we have selected eight EoSs from the paper~\cite{Somasundaram:2021ljr}: the Skyrme SLy4~\cite{Chabanat:1997un,Chabanat:1998} interaction which has been employed in several LVC papers~\cite{LIGOScientific:2017vwq}, SLy4$^*$ which is a modified version of SLy4 by changing $K_{\sym}$ from -120 MeV to 125 MeV (named nucleonic in Ref.~\cite{Somasundaram:2021ljr}), a set of three EoSs based on SLy4$^*$ with a first order phase transition (FOPT) occurring at $n_t=2.0n_\sat$, with the sound speed $c^2=2/3$ and three choices for the step density ($\delta n_t=1.15 n_\sat$ FOPT1, $\delta n_t=0.8 n_\sat$ FOPT2, $\delta n_t=0.5 n_\sat$ FOPT3), and finally, we have also considered three quarkyonic models~\cite{McLerran:2019} adapted to beta-equilibrated matter in compact stars~\cite{Margueron:2021} varying the parameter $\Lambda_\qyc$: qyc1 ($\Lambda_\qyc=332$~MeV), qyc2 ($\Lambda_\qyc=300$~MeV), qyc3 ($\Lambda_\qyc=275$~MeV).
The motivation for SLy4$^*$ is to generate a nucleonic model compatible with GW170817 as well as the NICER observations for the massive NS PSR J0740+6620~\cite{Somasundaram:2021ljr}.
In Fig.~\ref{fig:rayon_lambda}(a), the mass-radius relations for the eight EoSs employed in this analysis are shown: SLy4 and SLy4$^*$ (solid lines) are smooth, with SLy4 predicting systematically lower radii than SLy4$^*$, the three EoSs with a FOPT (green dashed lines) present a kink (reflecting the first order character of the transition) from where the radius sharply reduces as a function of the mass, and finally the three qyc models (dotted lines) that have a smooth mass-radius relation as a consequence of the transition between nucleonic and quark matter being a cross-over. The latter modeling predict an increase of the radius for mass in the range observed by GW170817.

We show in Fig.~\ref{fig:rayon_lambda}(b) the tidal deformability-mass relation, where the tidal deformability is defined as
\begin{equation}
\Lambda = \frac{2}{3G} k_2 R^5 \, ,
\end{equation}
with $k_2$ the tidal Love number and $R$ the NS radius, see for instance Ref.~\cite{Hinderer:2007mb} for more details. The more compact the NS, and thus massive, the weaker the tidal deformability, and the extreme case of zero deformability describes a black hole. By fixing, in what follows, the EoS in the GW signal analyses, we are in practice imposing that the relation between $\Lambda$ and the mass $m$ follows the relation given in Fig.~\ref{fig:rayon_lambda}(b). The vertical bar shows the effective tidal deformability $\tilde{\Lambda}$ obtained from GW170817~\cite{LIGOScientific:2018hze}, illustrating that our choice of EoSs widely explores the observational data.

\subsection{Comparison of EoS-blind analysis of GW170817 with analyses based on several EoSs}
\label{sec:gw170817}

\begin{figure}[t!]
\centering
\includegraphics[width=0.7\textwidth]{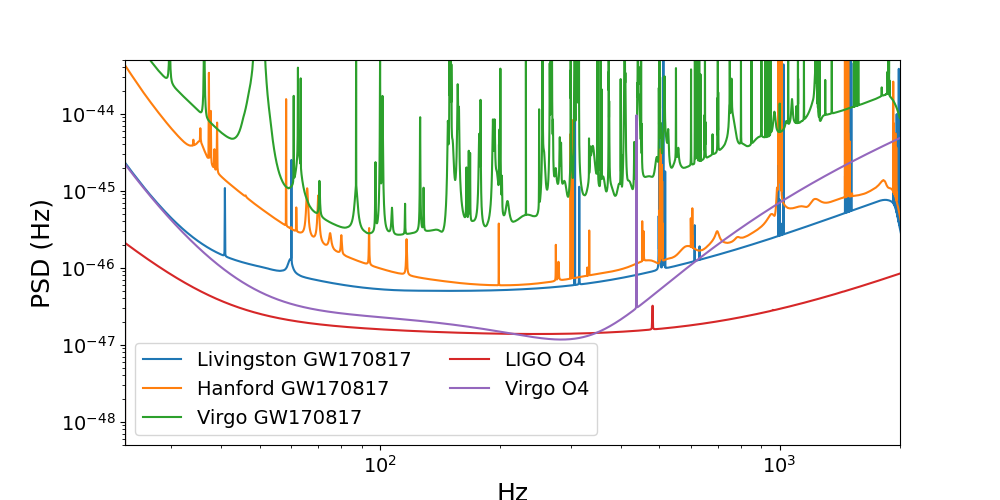}
\caption{Representative power spectral density of the three detectors' strain sensitivity. The noise curves during event GW170817 of run O2 are shown in green, orange and blue~\cite{LIGOScientific:2017vwq}. The design noise curves for O4 are shown in red and purple and are used to perform the O4 simulations~\cite{psd_ad}.} 
\label{fig:psd}
\end{figure}

The data $d$ ($d=h+n$) associated to a detection consists of the signal $h$ and the noise $n$, which is modeled by the power spectrum density (PSD). For GW170817, the PSDs characterizing the detectors LIGO Livingston, LIGO Hanford and Virgo are respectively plotted in blue, orange and green in Fig.~\ref{fig:psd}. The GW constituting the signal can be modeled by the IMRPhenomPv2\_NRTidal approximant~\cite{Dietrich:2017aum, Dietrich:2018uni, Dietrich:2019kaq}. This model is based on a pN development to which a high frequency phase evolution fit combining both an analytical EOB model~\cite{PhysRevLett.114.161103} and a set of numerical relativity simulations~\cite{PhysRevD.96.121501, Dietrich:2018upm} to better model tidal effects, called NRTidal, have been added. The template waveform gives the strain $h(t)$ as a function of 17 intrinsic and extrinsic parameters. The extrinsic parameters are the location in the sky of the source, its distance, its polarization, its inclination angle and its phase at the coalescence time. The intrinsic parameters are the masses of the two objects, their spins and their tidal deformabilities. 
The waveform is governed by a combination of these parameters. For instance the fifth and sixth orders in the post-Newtonian expansion are controlled by the effective tidal deformability $\tilde{\Lambda}$ and its effective asymmetric part $\delta \tilde{\Lambda}$ defined as~\cite{Wade:2014vqa}
\begin{equation}
\begin{aligned}
%\tilde{\Lambda} = \frac{16}{13(m_1+m_2)^5} \left(( m_1+12m_2)m_1^4\lambda_1 + ( m_2+12m_1)m_2^4\lambda_2 \right) & \\
%\delta \tilde{\Lambda} = \frac{1}{1319(m_1+m_2)^6} [(1319m_1^2-7996m_1m_2-11005m_2^2)m_1^4 & \lambda_1 \\
% +(11005m_1^2-7996m_1m_2-1319m_2^2) & m_2^4\lambda_2 ]
\tilde{\Lambda} = \frac{8}{13}\Big[ &(1+7\eta-31\eta^2)(\Lambda_1+\Lambda_2)+\sqrt{1-4\eta}(1+9\eta-11\eta^2)(\Lambda_1-\Lambda_2)\Big] \\
\delta \tilde{\Lambda} = \frac{1}{2} \Big[&\sqrt{1-4\eta}\left(1-\frac{13272}{1319}\eta+\frac{8944}{1319}\eta^2\right)(\Lambda_1+\Lambda_2)\\
& + \left(1-\frac{15910}{1319}\eta+\frac{32850}{1319}\eta^2 + \frac{3380}{1319}\eta^3\right)(\Lambda_1-\Lambda_2) \Big]
\end{aligned}
\label{eq:tilde}
\end{equation}
with $\eta=m_1m_2/m_\mathrm{tot}^2$, $m_\mathrm{tot}=m_1+m_2$, with $m_1$ and $m_2$ being the masses of the two NSs with $m_1\geq m_2$.

The determination of the 17 extrinsic and intrinsic parameters $\theta$ is performed by a Bayesian analysis, where the probability density function (PDF) $p(\theta | d, \mathcal{M}_A)$ is defined from the likelihood $\mathcal{L}(d|\theta, \mathcal{M}_A)$, the prior $\pi(\theta| \mathcal{M}_A)$, and the evidence $\mathcal{Z}(d | \mathcal{M}_A)$ according to the following formula~\cite{Thrane:2018qnx}:
\begin{equation}
    p(\theta | d, \mathcal{M}_A) = \frac{\mathcal{L}(d|\theta, \mathcal{M}_A)\pi(\theta| \mathcal{M}_A)}{\mathcal{Z}(d | \mathcal{M}_A)}\,.
\end{equation}
The symbol $\mathcal{M}_A$ represents the list of model parameters, including the waveform and the EoS parameters. Note that in the case where we haven't considered the EoS contribution (EoS blind analysis), there are therefore 17 parameters in total. In the case where an EoS is considered, there are only 15 parameters because the tidal deformability of a NS becomes a function of its mass.

\begin{table}[tb]
\centering
\begin{tabular}{|l|l|}
\hline{\bf Parameters} $\theta$ & {\bf Priors} $\pi(\theta)$ \\
\hline Chirp mass $\mathcal{M}=(m_1m_2)^{3/5}\, m_{tot}^{-1/5}$ &  \texttt{Uniform} $\mathcal{M}\in [$1.18$,$1.21$]M_{\odot}$ \\
\hline Mass ratio $q=m_2/m_1$ & \texttt{Uniform} $q\in [0.125,1]$ \\
\hline Spins $a_1$, $a_2$, $\theta_1$, $\theta_2$, $\phi_{12}$, $\phi_{jl}$ &  \texttt{Uniform} $a_1,a_2\in [0,0.05]$; \texttt{Sin} $\theta_1,\theta_2 \in [0,\pi]$; \texttt{Uniform} $\phi_{12},\phi_{jl} \in [0,2\pi]$ \\
\hline Sky localization $\alpha$, $\delta$ & $\alpha=3.446$ rad; $\delta=-0.408$ rad \\
\hline Luminosity distance $d_L$ & \texttt{Square PowerLaw} $d_L\in[0,80]$Mpc\\
\hline Orbital plane $\Psi$, $\theta_{jn}$ & \texttt{Uniform} $\Psi\in [0,\pi]$; \texttt{Sin} $\theta_{jn}\in [0,\pi]$ \\
\hline Coalescence phase $\phi_c$ & \texttt{Uniform} $\phi_c\in [0,2\pi]$\\
\hline Geocenter time $t_c$ & \texttt{Uniform} $t_c\in[\text{trigger-time}-0.1,\text{trigger-time}+0.1]s $ \\
\hline Tidal deformabilities $\Lambda_1$, $\Lambda_2$ & \texttt{Uniform} $\Lambda_1, \Lambda_2 \in [0, 5000]$\\ 
\hline
\end{tabular}
\caption{Priors used for the Bayesian analysis.}
\label{tab::prior}
\end{table}

The Bayesian analysis of GW170817 is performed by using Parallel Bilby v1.1.0 \cite{pbilby_paper, bilby_paper}, a parallelized Bayesian inference Python package, and Dynesty v1.0.1 \cite{dynesty_paper, skilling2004, skilling2006}, a nested sampler. For production runs, the priors used are given in Table~\ref{tab::prior}. In order to speed-up the numerical calculations, the right ascension and declination are fixed from the observation of the EM counterpart. We use the default parameters recommended by the LVC with the phase and distance marginalization, nlive=1000 (number of live points), nact=10 (to ensure that the minimum p-value in pp tests~\cite{doi:10.1198/106186006X136976} is above 1/15) and n-parallel=4 (number of independent jobs per event to improve the smoothness of results). For BNS systems with the range of chirp mass considered, the signal duration $T$ of the leading order in pN expansion of the inspiral starting at a frequency $f_0\simeq 20$~Hz is:
\begin{equation}
T = \frac{5}{256 f_0 \eta \pi^{8/3}}\left(\frac{f_0 G m_\mathrm{tot}}{c^3} \right)^{-5/3} \simeq 165~ \text{s} \,,
\end{equation}
where $G$ is the Newton's constant $c$ the speed of light. In the following we use therefore a signal duration of $192$~s. In the case of BNS systems the computational power needed for a Bayesian inference analyzing described above is quite expensive, taking about one day with the power of a cluster with 8 processors Intel Cascade Lake 6248 ($8\times 20$ cores at 2.5 GHz).

\begin{figure}[t!]
\begin{subfigure}{0.49\textwidth}
    \includegraphics[width=\textwidth, trim = {5 0 30 0}, clip]{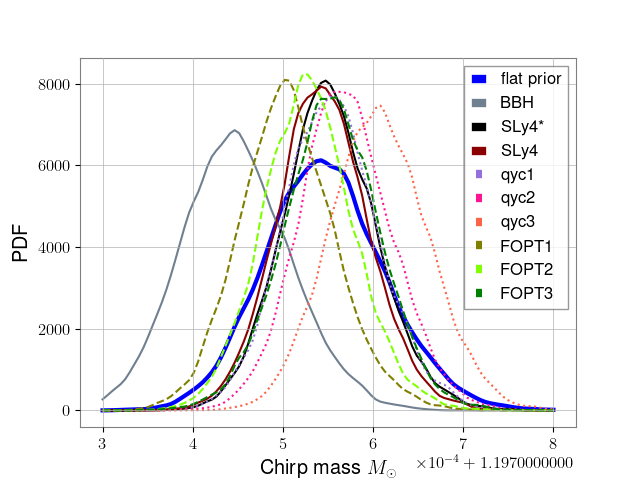}
    \caption{}
    \label{afig:gw170817_MQ}
\end{subfigure}
\begin{subfigure}{0.49\textwidth}
    \includegraphics[width=\textwidth, trim = {5 0 30 0}, clip]{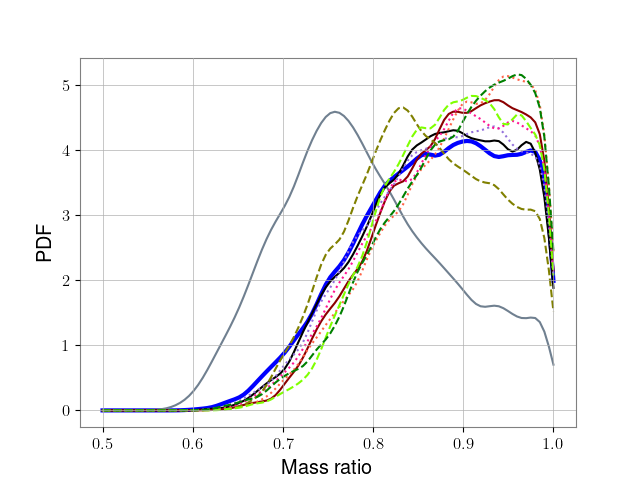}
    \caption{}
    \label{bfig:gw170817_MQ}
\end{subfigure}
\caption{The PDF of the chirp mass $\mathcal{M}$ (a) and the mass ratio $q$ (b) obtained from the analysis of the GW170817 observed data for various EoSs considered in this analysis (see the legend for more details). The \textsl{flat prior} blue curve is obtained assuming a flat prior in $\mathcal{M}$ and $q$ without assuming a given EoS as in the original LVC analysis~\cite{LIGOScientific:2017vwq} (see text for more details) and the \textsl{BBH} curves assumes $\Lambda_1=\Lambda_2=0$.} 
\label{fig:gw170817_MQ}
\end{figure}

A full Bayesian PDF is constructed from the comparison of the observed GW170817 GW signal and the modelled one, as previously explained. The 1D marginal distributions of $\mathcal{M}$ and $q$ are shown in Fig.~\ref{fig:gw170817_MQ} for different approaches for the EoS: the {\it flat prior} analysis refers to a choice of a uniform prior for the tidal deformabilities, independently of the choice of a given EoS, while the other curves assume a choice in the EoS. The different EoSs produce a dispersion of the position of the peak in the chirp mass PDF (see Fig.~\ref{fig:gw170817_MQ}, (a)) and this dispersion is slightly larger than the width of the flat prior PDF. In all cases, the chirp mass is however very well bounded between $1.1973$~$M_{\odot}$ and $1.1978$~$M_{\odot}$. For the BBH case the associated PDF is peaked at $1.1975\pm0.0002$~$M_{\odot}$ whereas for the other cases (with matter described by our sample of EoSs), the FOPT1 and the qyc3 models represent the two extreme PDFs for the mass dispersion. Note that these two EoSs are also the ones constraining the radius of a $1.4 M_\odot$ NS ($R_{1.4}$) to be between 11.4 and 13.3~km, see Fig.~\ref{fig:rayon_lambda}(a). The mass ratio PDF (see Fig.~\ref{fig:gw170817_MQ}, (b)) for the BNS points towards a 90\% confidence region above $0.67$ with a mean value of $0.9$. There are also some differences between different choices of the EoS concerning the position of the mean. For all the EoSs explored in this study, we find that the PDF for the spins, the PDF for the luminosity distance and the PDF for the orientation of the coalescence plane of the source are statistically compatible. However, the geocentric time is much better determined by fixing an EoS compared to the usual {\it flat prior} analysis. As previously explained, the two angles of the sky location have been fixed.

It is interesting to remark that the BBH and BNS PDF for the chirp mass and the mass ratio have a large overlap in Fig.~\ref{fig:gw170817_MQ}. The BBH PDF is peaked at a lower value, $q=0.72^{+0.15}_{-0.06}$, which makes it still compatible with the BNS PDF. This indicates that a GW signal alone can not help distinguishing between a BBH or a BNS origin, and that one needs additional information for this purpose, for instance on the true mass distribution or on the electromagnetic counter parts of the gravitational wave, see for instance Refs.~\cite{PhysRevD.101.103008, PhysRevD.104.084006, PhysRevD.105.064063}.

\begin{figure}[t!]
\centering
\includegraphics[width=0.55\textwidth]{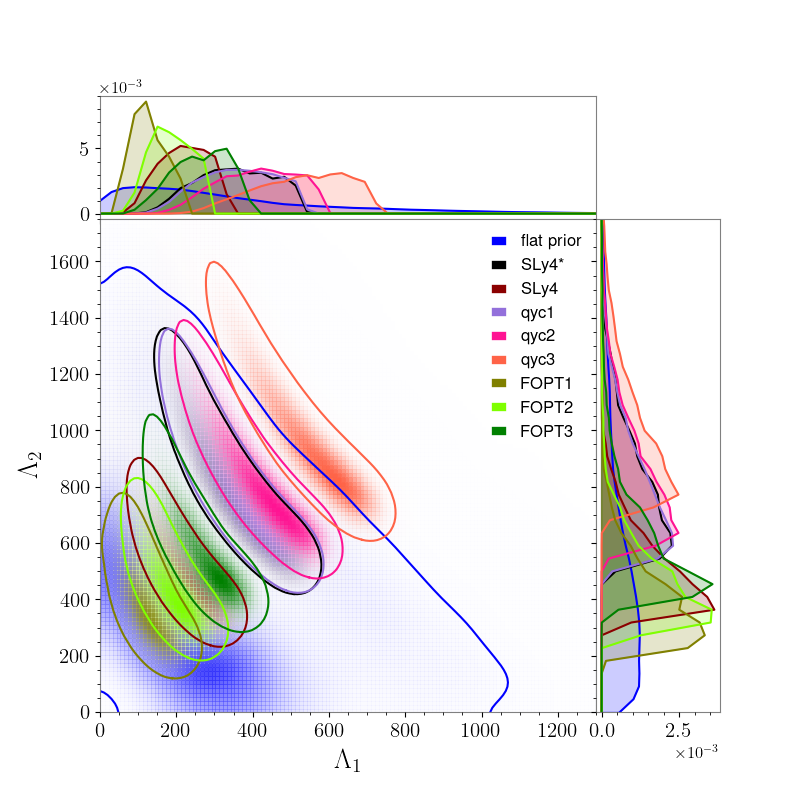}
\caption{Analyses of GW170817 signal in $\Lambda_1$-$\Lambda_2$ plane based on different assumptions identical to the ones shown in Fig.~\ref{fig:gw170817_MQ} for the BNS (BBH is excluded here). The 1-D projections of the Bayesian distributions are shown in the upper and right additional insets.}
\label{fig:gw170817_lambda1_lambda2}
\end{figure}

The tidal deformability $\Lambda_i$ is a function of the mass $m_i$ when a given EoS is fixed in the Bayesian analysis, while the \textsl{flat prior} approach does not assume a relation between these intrinsic parameters. This has an impact on the values for $\Lambda_1$ and $\Lambda_2$ explored by the Bayesian analysis, as shown in Fig.~\ref{fig:gw170817_lambda1_lambda2}. The contours associated to a given EoS are systematically smaller than the \textsl{flat prior} one. The values for $\Lambda_1$ and $\Lambda_2$ are also different and depend on the considered EoS. The variation of the lengths of the EoS distributions in the direction $\Lambda_1=-\Lambda_2$ reflects the uncertainty in the mass ratio $q$, while the thickness of the contours in the symmetric direction is mainly due to the very small uncertainty in the chirp mass $\mathcal{M}$. 
It can be noted that, following the $\Lambda_1 = \Lambda_2$ line, the contours are crossed by increasing compactness of the star approximately measured at 1.4$M_\odot$, as already pointed out in the original LVC paper~\cite{LIGOScientific:2017vwq}. The contours associated to the SLy4$^*$ and qyc1 EoSs are very well overlapping, reflecting that their tidal deformabilities for masses lower than 1.6$M_\odot$ (the upper mass explored by GW170817) are almost identical, see Fig.~\ref{fig:rayon_lambda}.

\begin{figure}[t!]
\centering
\includegraphics[width=0.6\textwidth]{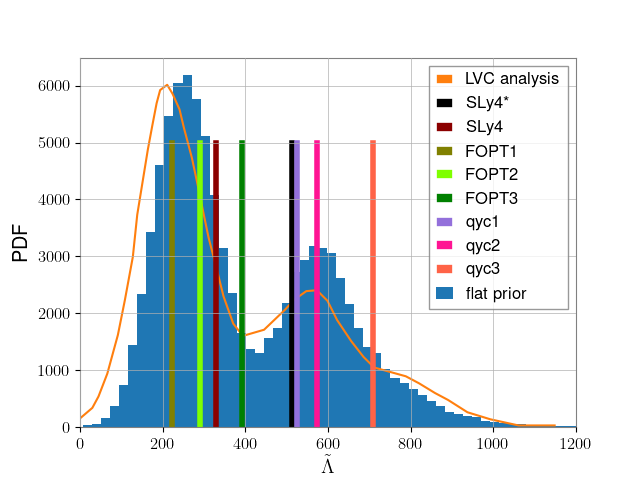}
\caption{The PDF of the effective tidal deformability $\tilde{\Lambda}$ obtained from the analysis of the observed data. Our analysis on a 192~s signal is compared to the original one~\cite{LIGOScientific:2017vwq} (solid orange curve) with a signal of 128~s. Despite the small difference, this comparison shows that the two analyses are quite compatible. The vertical bars represent the effective tidal deformability obtained from the eight EoSs we consider in this analysis and assuming the PDF of the masses.} 
\label{fig:gw170817_lambda}
\end{figure}

In the GW analysis, the most important parameter connected to the EoS is the effective tidal deformability $\tilde{\Lambda}$ intervening at order 5pN in the waveform. When an EoS is considered, this value is calculated by Eq.~\eqref{eq:tilde} from the PDF of the masses: $\tilde{\Lambda}=\tilde{\Lambda}(m_1,m_2,\Lambda_1(m_1),\Lambda_2(m_2))$. The sharp prediction for $\tilde{\Lambda}$ has a 90\% credible level of about 22 while its direct measurement without consideration of an EoS yields a broader PDF: $\tilde{\Lambda}=337^{+440}_{-180}$. This PDF is shown in Fig.~\ref{fig:gw170817_lambda} assuming a \textsl{flat prior} in the intrinsic parameters $\tilde{\Lambda}$ and $\delta\tilde{\Lambda}$ and the average value of the PDF for each EoS in our set is represented by a vertical bar. We obtain a noticeable secondary peak in the PDF of $\tilde{\Lambda}$, similar to the one obtained in previous LVC analyses~\cite{LIGOScientific:2017vwq,LIGOScientific:2018hze} and which we represent in orange in the same figure. The origin of this peak is not fully understood~\cite{LIGOScientific:2017vwq, LIGOScientific:2018hze}, and it may contribute to increase the uncertainty in $\tilde{\Lambda}$.
The relative size of the secondary peak is varying with the model used for the waveform, see for instance the comparison shown in Fig.11 from Ref.~\cite{LIGOScientific:2018hze}: The analysis based on the Taylor-F2 model~\cite{PhysRevD.85.123007}, which is a purely analytic PN model, has a secondary peak which is reduced compared to the analyses based on other waveform models. This difference might be one of the reasons explaining the values for the effective tidal deformabilities extracted by different authors: 
from~\cite{PhysRevLett.121.091102} $\tilde{\Lambda}=222^{+420}_{-138}$ (by using only Taylor-F2), while from the LVC~\cite{LIGOScientific:2018hze} $\tilde{\Lambda}=300^{+420}_{-230}$ (by averaging over several waveform models). The Taylor-F2 model is however known to miss some important ingredients in its parametrisation while other models are more complete, such as for instance the model IMRPhenomPv2\_NRTidal that we have considered here.
% nous $\tilde{\Lambda}=337^{+440}_{-180}$ (low spin prior)
%De at al obtained $\tilde{\Lambda}=222^{+420}_{-138}$
%Abbott:2017: $\tilde{\Lambda}\leq 800$ (low spin prior)
%Abbott:2019 (PRX9): $\tilde{\Lambda}=300^{+420}_{-230}$ (low spin prior)

When an EoS is fixed during the analysis, both the PDF of $\tilde{\Lambda}$ and geocentric time are much better determined than during the standard analysis. In appendix \ref{app:time}, we discuss the correlation between the double peak in $\tilde{\Lambda}$ and the double peak distribution in the geocentric time. This correlation points towards an explanation of the double peak in the GW signal observed in GW170817: the difficulty to properly assign a geocentric time to the arrival of the signal. This could be due to the presence of noise, as we will illustrate in the next section, linking the uncertainty in the geocentric time to the low signal to noise ratio (SNR).

\subsection{Bayes factor associated to different analyses with a given EoS}

\begin{table}[tb]
\centering
\begin{tabular}{|c|c|c|}
\hline $|\ln \mathcal{B}_{AB}|$ & Probability & \\
\hline  $<1$    &   $<0.731$   & Inconclusive \\
\hline  $2.5$ & $0.924$ & Moderate evidence \\
\hline  $5$ & $0.993$ & Strong evidence \\
\hline
\end{tabular}
\caption{Jeffrey's scale standard values used to compare two competing models using the Bayes factor~\cite{Jeffreys}.}
\label{tab::bayes_factor}
\end{table}

To compare the ability of two models, $\mathcal{M}_A$ and $\mathcal{M}_B$, to describe the same data we use the Odds factor $\mathcal{O}_{AB}$, defined by the ratio between $p(\mathcal{M}_A |d)$, the probability of model $\mathcal{M}_A$ given $d$, and $p(\mathcal{M}_B |d)$, the probability of model $\mathcal{M}_B$ given $d$. According to~\cite{Thrane:2018qnx}, this factor is equal to:
\begin{equation}
\mathcal{O}_{AB} = \frac{p(\mathcal{M}_A |d)}{p(\mathcal{M}_B |d)} = \frac{p(\mathcal{M}_A)}{p(\mathcal{M}_B)} \frac{\mathcal{Z}_A}{\mathcal{Z}_B}\,,
\label{Bayes_factor}
\end{equation}
with $p(\mathcal{M}_{A/B})$ being the prior on the model. If no model is a priori preferred, which will be our case even for the BBH model, the Odds factor is directly equal to the Bayes factor defined by $\mathcal{B}_{AB}=\mathcal{Z}_A/\mathcal{Z}_B$. The meaning of the Bayes factor in terms of an evidence is given in table~\ref{tab::bayes_factor} presenting the Jeffrey's scale of empirical evidence~\cite{Jeffreys}. A value of $2.5$ shows a preference for a model, while a value of $5$ represents a strong evidence. 

Each of the models used to analyze the GW170817 event is defined by considering an EoS. All these models fit the data well even if they present some differences in the PDF as discussed in the previous section. We can use the the Bayes factor to rank these models, and we show the results in Table~\ref{tab::BF} with respect to the SLy4 EoS, chosen as a reference model. As expected $\mathcal{B}_{AB}$ is equal to zero in the case where $\mathcal{M}_A$ is also SLy4. The modified version of SLy4, the three FOPT models with a first-order phase transition, and the three quarkyonic models all have Bayes factors between -2.5 and 2.5. Even with a $40$Mpc source, the SNR of the GW170817 event is not large enough to discard any of the EoSs considered in~\ref{EOS}. We will have to wait for another exceptional event with a better resolution to exclude one of the two families. The modeling of the emitting source by two black holes with considered zero tidal deformabilities is moderately disfavoured compared to the SLy4 EoS because its Bayes factor is lower than $-2.5$.    

The value of the tidal deformability determined from the source masses and using an EoS is contained within 90\% of the confidence level of the tidal deformability measured by the Bayesian analysis using a uniform prior on $\Lambda_1$ and $\Lambda_2$ (see Fig.~\ref{fig:gw170817_lambda1_lambda2} and Fig.~\ref{fig:gw170817_lambda}). The calculation of the Bayes factor to rank different models is costly from the computational point of view. We have therefore introduced an approximation to this calculation, which is described in detail in appendix~\ref{app:BF}, allowing to quickly determine the Bayes factors for all EoS analyses with respect to the {\it flat prior} one. This method is particularly relevant for moderate evidence.

\begin{table}[tb]
\centering
\begin{tabular}{|l|c|c|c|c|c|c|c|c|c|}
\hline EoS &          BBH & \hspace{0.1cm}FOPT1\hspace{0.1cm} & \hspace{0.1cm}FOPT2\hspace{0.1cm} & \hspace{0.1cm}FOPT3\hspace{0.1cm} & \hspace{0.1cm}SLy4*\hspace{0.1cm} & \hspace{0.1cm}SLy4\hspace{0.1cm} & qyc1 & qyc2 & qyc3 \\
\hline  Bayes factor\hspace{0.1cm} &  $-2.98$ & $0.40$ & $0.27$ & $-1.19$ & $-1.03$ & 0. & $-1.21$ & $-0.85$ & $-2.13$  \\
\hline
\end{tabular}
\caption{Bayes factor with respect to the SLy4 EoS for all models considered.}
\label{tab::BF}
\end{table} 

\section{Simulated data anticipating the O4 run and new detections of BNS mergers}

Since the O2 run, several technical improvements~\cite{aLIGO:2020wna} have been implemented in LIGO and Virgo in order to increase the sensibility of the facilities and to reach the design value for O4 shown in Figure~\ref{fig:psd} (the sensitivity is expected to be $5$ times better in O4 than in O2). In this section, we address the question of what would be the ability during O4 to discriminate between different EoSs in case of a possible repetition of an event like GW170817. We also perform this study as a function of the distance of the source.

We first estimate the probability of having a BNS merger as close as GW170817 i.e. at a distance of less than $40$~Mpc. From the BNS merger rate estimated in
Ref.~\cite{Mochkovitch:2021prz}, $\tau_\mathrm{BNS}=320^{+490}_{-240}$~Gpc${}^{-3}$y${}^{-1}$, one could deduce the average BNS event rate,
\begin{equation}
 R^{-1} = \left(\tau_\mathrm{BNS}\frac{4\pi}{3}D^3_{\text{\tiny GW170817}}\right)^{-1} = 12^{+36}_{-7}\text{yr}\,.
\end{equation}
Since O4 is expected to take data for about 1 year, we have only 1 out of 12 chances (considering the centroid) to detect an event like GW170817 during the next O4 run. The value is fairly low and does not even take into account a favorable orientation of the detectors, but it also suffers from a very large uncertainty. However, since the sensitivity will be improved in O4, compared to O2, a larger horizon will be accessible and thus the number of observed events is expected to be larger. It is not straightforward to estimate quantitatively how much of this larger space explored by O4 will contribute to improve the accuracy of the determination of the effective tidal deformability $\tilde{\Lambda}$, because one needs to also account for the evolution of the SNR, which is an increasing function of the distance.
In the following, we illustrate this point better and present quantitative results on the ability of LVK to determine the dense matter EoS in O4, compared to the present knowledge obtained from GW170817.

\subsection{Simulated data like GW170817 with the O4 PSD}

Despite the fact that the O4 run is not started yet, we can simulate an event by assuming the various source parameters (like tidal deformabilities, mass ratio, localisation and geometric parameters) on top of the expected O4 noise. In practice, the GW is generated by the use of the IMRPhenomPv2\_NRTidal approximant with parameters as close as possible to the ones of the GW170817 event. The chirp mass and mass ratio are chosen to be, respectively, $1.19755M_{\odot}$ and $0.95$ (see Fig.~\ref{fig:gw170817_MQ}), the spins are chosen to be almost zero ($|\chi|<0.01$) and the location in the sky is at first fixed and set to be identical to GW170817 with a distance of $40$~Mpc. The injected tidal deformability is calculated for two EoSs (SLy4 and qyc2) by using their relation with the injected masses, as in Fig.~\ref{fig:gw170817_lambda1_lambda2}. On top of this signal we add a realisation of the noise based on the PSDs for the O2 or O4 runs, as shown in Fig.~\ref{fig:psd}.

\begin{figure}[t]
\centering
\begin{subfigure}{0.49\textwidth}
    \includegraphics[width=\textwidth]{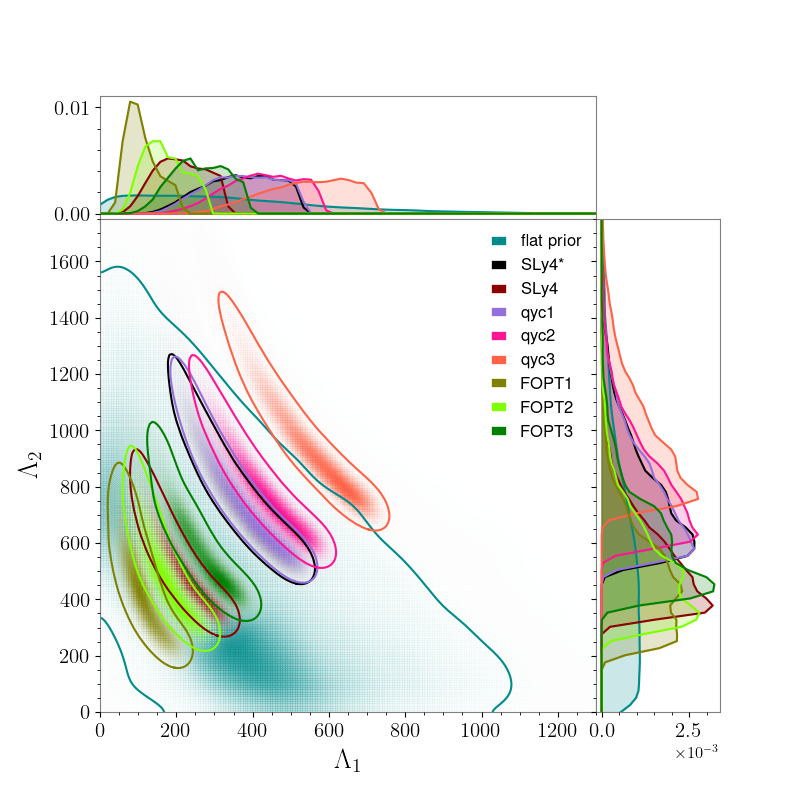}
    \caption{Injection with SLy4 and O2 PSD.}
    \label{afig:lambda1_lambda2}
\end{subfigure}
\begin{subfigure}{0.49\textwidth}
    \includegraphics[width=\textwidth]{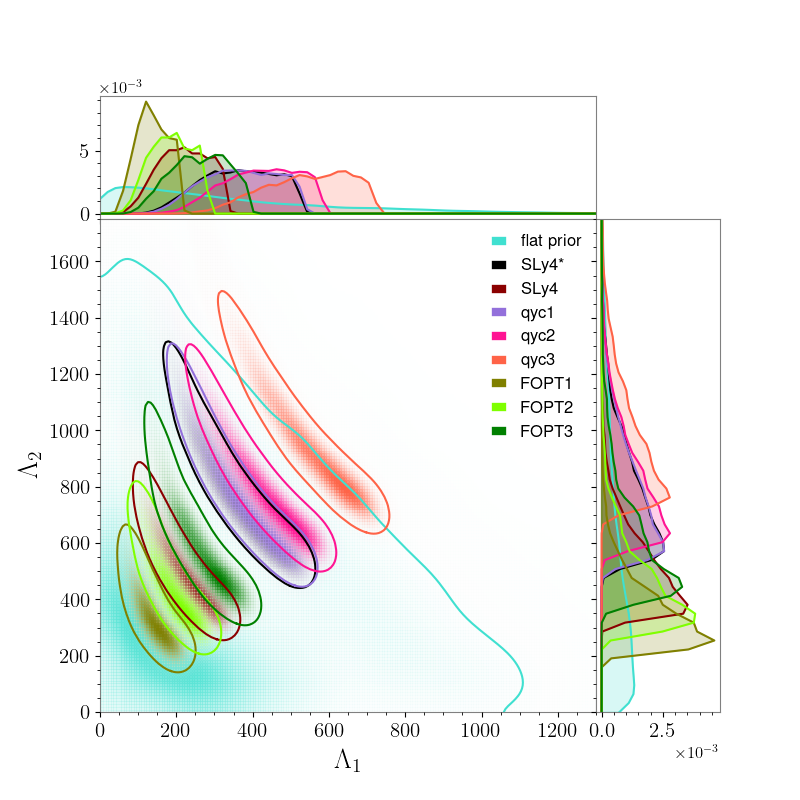}
    \caption{Injection with qyc2 and O2 PSD.}
    \label{cfig:lambda1_lambda2}
\end{subfigure}
\begin{subfigure}{0.49\textwidth}
    \includegraphics[width=\textwidth]{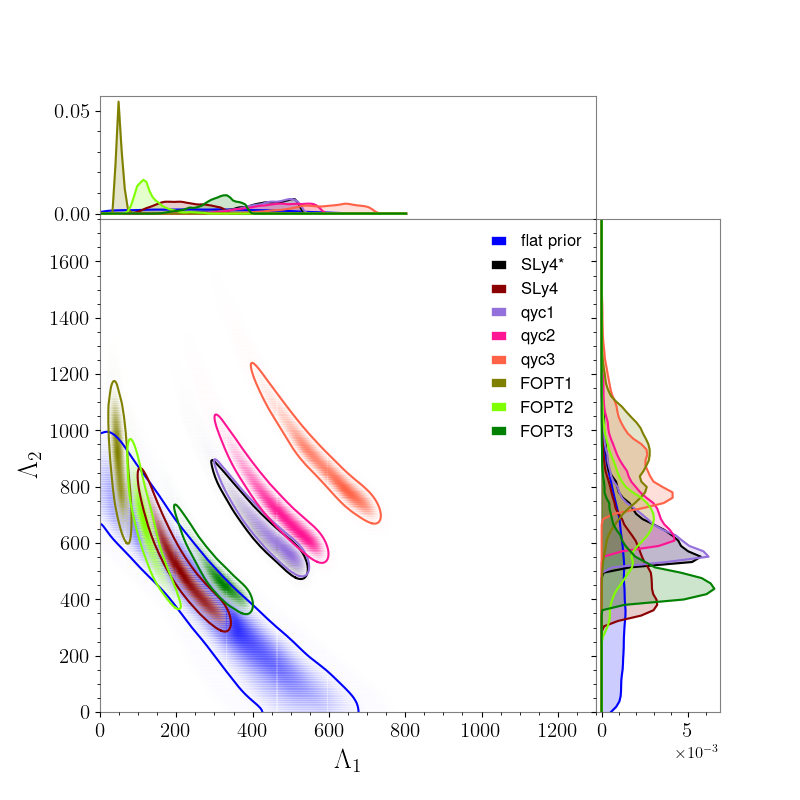}
    \caption{Injection with SLy4 and O4 PSD.}
    \label{bfig:lambda1_lambda2}
\end{subfigure}
\begin{subfigure}{0.49\textwidth}
    \includegraphics[width=\textwidth]{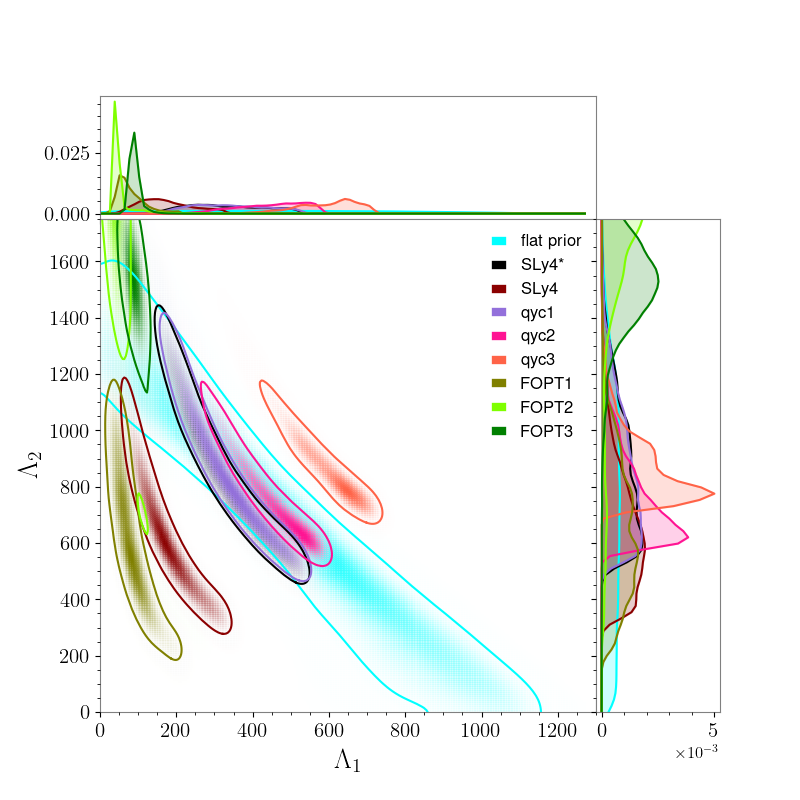}
    \caption{Injection with qyc2 and O4 PSD.}
    \label{dfig:lambda1_lambda2}
\end{subfigure}
\caption{$\Lambda_1$-$\Lambda_2$ posterior correlation obtained from the injection of a GW signal similar to GW170817.} 
\label{fig:lambda1_lambda2}
\end{figure}

\subsubsection{Impact of O4 on $\Lambda_1$-$\Lambda_2$ posterior correlation}
\label{sec:L1L2}

In the first analysis of GW170817 by the LVC, the measurement of the 2-dimensional contour of NS tidal deformabilities using a uniform prior is given in the paper~\cite{LIGOScientific:2017vwq}. An EoS fitting well the data should predict tidal deformabilities in this contour, as illustrated in Fig.~\ref{fig:gw170817_lambda1_lambda2}.    

In Fig.~\ref{fig:lambda1_lambda2} we show the $\Lambda_1$-$\Lambda_2$ posterior correlation obtained with the injected signal built on SLy4, panel (a), and qyc2, panel (b), and with the O2 PSD. These panels show that the small difference do not really allow for a separation between the different EoSs. In addition, the figures are both very similar to the one in Fig.~\ref{fig:gw170817_lambda1_lambda2}, which consolidates the realism of our approach in simulating real data.

Panels (c) and (d) in Fig.~\ref{fig:lambda1_lambda2} are similar to panels (a) and (b), but using the expected PSD for O4. In panel (c), 4 EoSs (SLy4$^*$, qyc1, qyc2, qyc3) do not predict a contour compatible with the {\it flat prior} analysis. The contours for FOPT1 and FOPT2 do not have the same orientation, and prefer an asymmetric system incompatible with the injected value to fit the data. In panel (d), where we use qyc2 for the injection, it is FOPT2 and FOPT3 that predict an asymmetric system. The mass ratio is less well measured with the FOPT1, SLy4, SLy4$^*$ and qyc1 hypothesis than with the {\it flat prior} analysis, which explains the very elongated contours in the diagram. With the O4 PSD only the use of the injected EoS predicts a zone perfectly compatible with the {\it flat prior} analysis. The clear differences in the posteriors of other analyses illustrate a better ability, in O4, to select among the candidate EoSs.     

\subsubsection{Impact of O4 on $\tilde{\Lambda}$ PDF}
\label{sec:LT}

\begin{figure}[t!]
\centering
\centering
\begin{subfigure}{0.49\textwidth}
    \includegraphics[width=\textwidth]{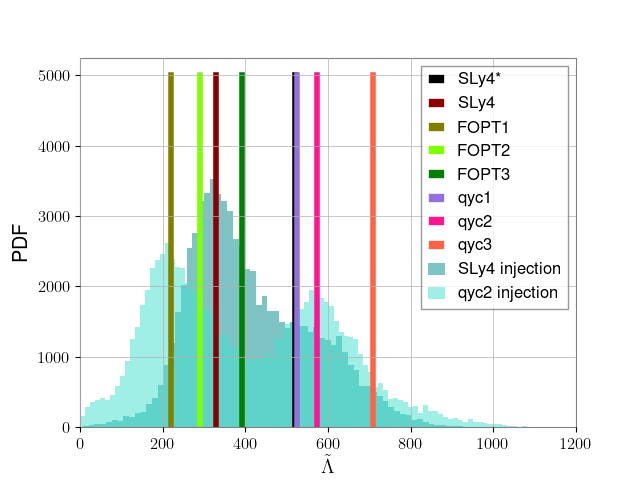}
    \caption{Injected data with the O2 PSD.}
    \label{afig:effectivelambda}
\end{subfigure}
\begin{subfigure}{0.49\textwidth}
    \includegraphics[width=\textwidth]{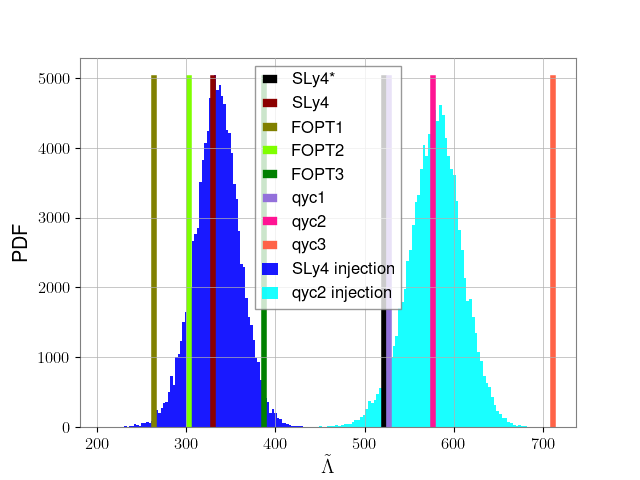}
    \caption{Injected data with the O4 PSD.}
    \label{bfig:effectivelambda}
\end{subfigure}
\caption{The $\tilde{\Lambda}$ PDF obtained from the injection of a GW signal similar to GW170817 and assuming SLy4 or qyc2 EoS, to be compared with Fig.~\ref{fig:gw170817_lambda} showing the real data. The vertical bars represent the mean value of the effective tidal deformability obtained from the eight EoSs.}
\label{fig:effectivelambda}
\end{figure}

As previously discussed, the GW measures the effective tidal deformability much better than the individual NS deformabilities. The resulting posterior $\tilde{\Lambda}$ PDF is shown in Fig.~\ref{fig:effectivelambda}, for the same cases shown in Fig.~\ref{fig:lambda1_lambda2}: the O2 PSD is shown in panel a while the O4 PSD in panel b. With the O2 PSD, the 90$\%$ confidence interval is $\tilde{\Lambda}=383^{+400}_{-163}$ and $\tilde{\Lambda}=370^{+380}_{-253}$ for injected data with SLy4 and qyc2 respectively. For the O4 PSD, these values become $\tilde{\Lambda}=335^{+41}_{-43}$ and $\tilde{\Lambda}=580^{+48}_{-53}$. The 90\% confidence level decreases from about $600$ to $100$. So, if an event similar to GW170817 is observed during O4, the effective tidal deformability is expected to be determined with a precision about six times better. With the O2 PSD, both distributions predict mutually compatible $\tilde{\Lambda}$ values and contain all the EoSs considered represented by the vertical bars in Fig.~\ref{fig:effectivelambda}. As with the real data from GW170817, none of our EoSs can be disfavoured. With O4, the two $\tilde{\Lambda}$ PDF are symmetric and can be approximated by Gaussian distributions with a standard deviation of $26$ for the injection with SLy4 and $30$ for qyc2. Moreover, these PDFs do not overlap and can exclude some of the EoSs considered here.

\begin{figure}[t!]
\includegraphics[width=0.70\textwidth, trim = {0 52 0 0}, clip]{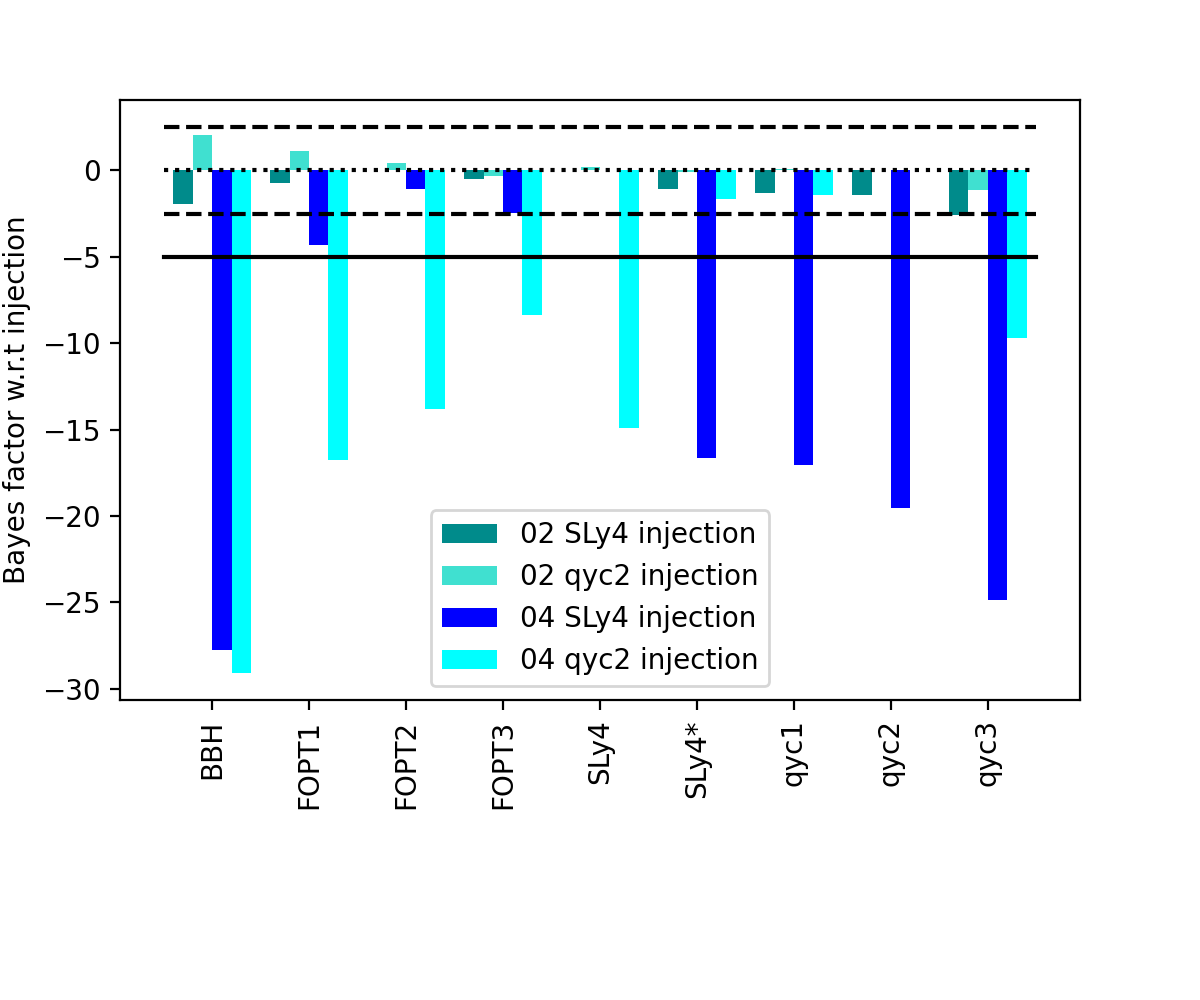}
\caption{Bayes factor calculated by Bayesian inference with respect to the EoS considered to create the injected signal (green bars for O2 and blue bars for O4). The dashed black horizontal line is the moderated evidence limit and the black horizontal line is the strong evidence limit, see Jeffrey's scale in Table~\ref{tab::bayes_factor}.}
\label{fig:histo}
\end{figure}

More quantitative statements about the ability to constrain the EoS may also be given by Bayesian factors. They are shown in Fig.~\ref{fig:histo} for O2 injections (green bars) and O4 injections (blue bars) with the simulated signal based on the SLy4 (dark color) or qyc2 (light color) EoS. The figure shows that the O2 injections are not able to distinguish between the different EoSs, while the O4 injections are clearly more selective. For instance, injecting SLy4 signal excludes SLy4$^*$ as well as qyc1, qyc2 and qyc3, while injecting qyc2 clearly excludes qyc3 as well as SLy4, FOPT1, FOPT2 and FOPT3. The two injected EoSs also exclude the BBH hypothesis.\\

\subsubsection{Effect of the NS masses}

These results are to be put in perspective with respect to the mass of the NSs. A more massive NS will give a smaller radius but also a smaller tidal deformability. For a NS of $1.9$~$M_{\odot}$, the tidal deformability is respectively $16$, $23$, $24$, $41$, $54$, $64$, $83$, $117$, for FOPT1, SLy4$^*$, FOPT2, FOPT3, nucleonic, qyc1, qyc2 and qyc3, that is to say a total variation of $100$ between the two extremes. This value is approximately the size of the 90\% confidence level region obtained with O4 in the previous section. The {\it flat prior} analysis of an injection with masses of $m_1=1.9$~$M_{\odot}$ and $m_2=1.8$~$M_{\odot}$ ($M=1.612$~$M_{\odot}$, $q=0.95$) gives two overlapping PDF. At such high masses, less favored by observations ~\cite{Lattimer:2012nd, Romani_2022, doi:10.1146/annurev-astro-081915-023322}, the distinction between different EoSs is much more difficult.

\subsection{Impact of the noise realisation from O4 PSD}
\label{sec:noise}

\begin{figure}[t!]
\begin{subfigure}{0.49\textwidth}
    \includegraphics[width=\textwidth]{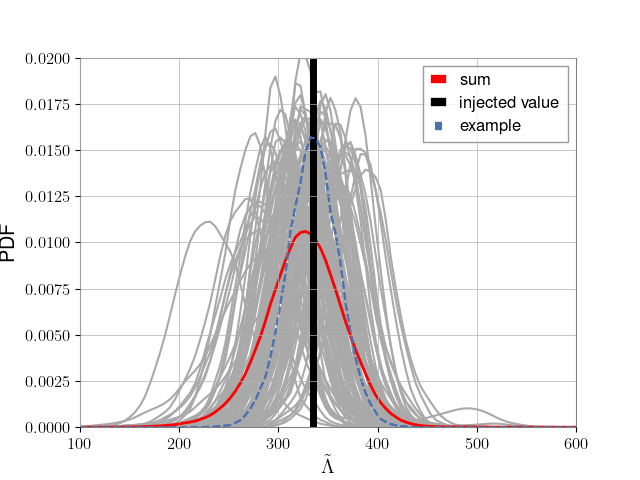}
    \caption{}
    \label{afig:120_seed}
\end{subfigure}
\begin{subfigure}{0.49\textwidth}
    \includegraphics[width=\textwidth]{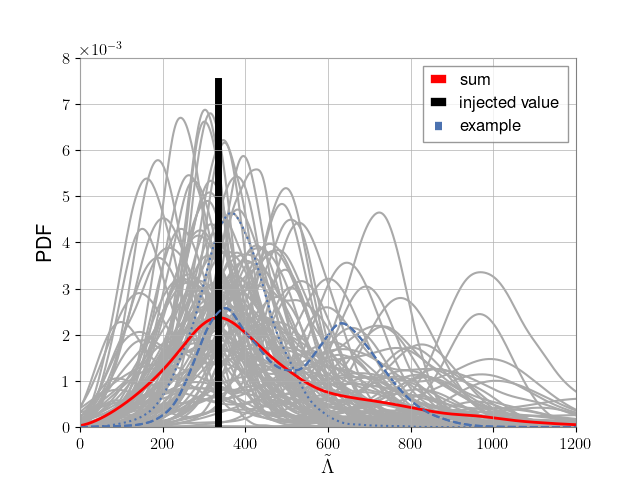}
    \caption{}
    \label{bfig:40_seed}
\end{subfigure}
\caption{Posteriors on $\tilde{\Lambda}$ from the same event employing different noise realisations with the O4 PSD for a source at 40 Mpc (panel a) and 120 Mpc (panel b). In blue we show some examples of the PDF, featuring a one or a two-peak structures, and in red the result after the normalized sum over the 100 different realisations of the noise.}
\label{fig:noise}
\end{figure}

In all previous analyses, we have used a particular realisation of the noise, given the PSD, for our predictions. The effect of changing the noise realisation by employing different seeds in the randomisation of the Gaussian noise is shown in Fig.~\ref{fig:noise}, in terms of reconstructed $\tilde{\Lambda}$. For each of the 100 noise realisations showing in gray in Fig.~\ref{fig:noise}, we have evaluated their chance to reproduce the injected value of $\tilde{\Lambda}$. We found that in 88 cases at 40~Mpc and in 91 cases at 120~Mpc, the injected signal was found within the 90\% credible interval of the PDF. The noise realisation plays an important role in the reconstruction of the signal and in 10\% of the cases it is expected that the injected value is not in the $90$\% confidence level region.   
The red curves in Fig.~\ref{fig:noise} show the average posterior probability density functions, obtained as the normalized sum of all curves in gray.

The realisation of the noise used in Fig.~\ref{fig:effectivelambda}(b) using the SLy4 EoS for injection is plotted as a blue dashed curve in Fig.~\ref{fig:noise}(a). This realisation gives an average value of the $\tilde{\Lambda}$ PDF close to the injected value with a region at 90\% confidence level equal to $84$. This value is slightly lower than what given by using the  average curve, in red, which is 93. We had thus fallen into a rather favorable case. For the case of farther distances, the average posterior is almost centered on the injected value with a region at 90\% confidence level of $130$. This is an increase of about $40$\% compared to the average value of $93$.    

With the signal injected at 40~Mpc using the O4 PSD, the $\tilde{\Lambda}$ PDF always has a Gaussian shape (see Fig.~\ref{fig:noise}(a)). When injecting at 120~Mpc, we obtain different shapes depending on the realisation of the noise. For example, in Fig.~\ref{fig:noise}(b), the blue dashed curve has a double peak while the blue dotted curve has a single peak. This means that at larger distances the SNR is not sufficient to obtain an accurate measurement of $\tilde{\Lambda}$ and, therefore, the measurement will depend more strongly on the realisation of the noise. We notice that the double-peak structure of the posterior is also present when analyzing real data from GW170817 and when analyzing simulated data with qyc2 at 40~Mpc with a noise realisation using the O2 PSD (see light green histogram in Fig.~\ref{fig:effectivelambda}(a)). Our observation hints towards a non-physical effect, rather induced by the noise level. 

\begin{figure}[t!]
\centering
\includegraphics[width=0.7\textwidth]{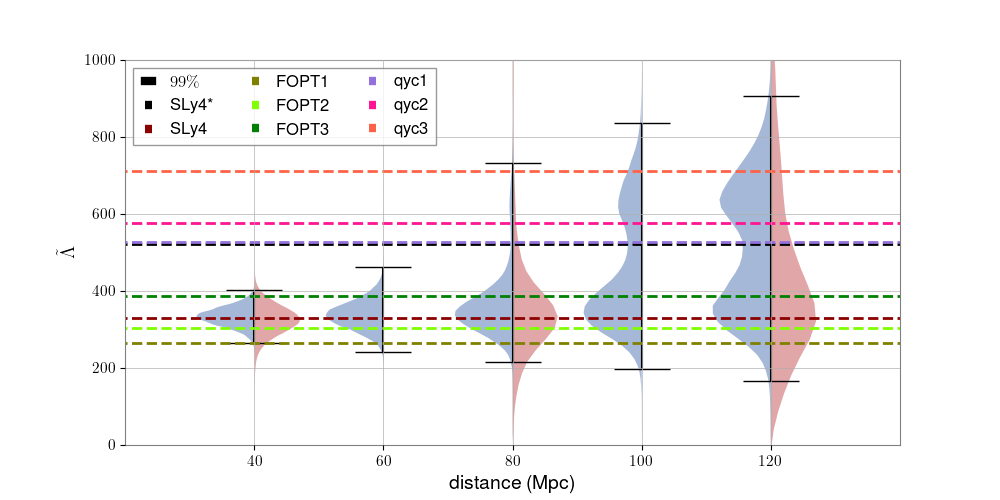}
\includegraphics[width=0.7\textwidth]{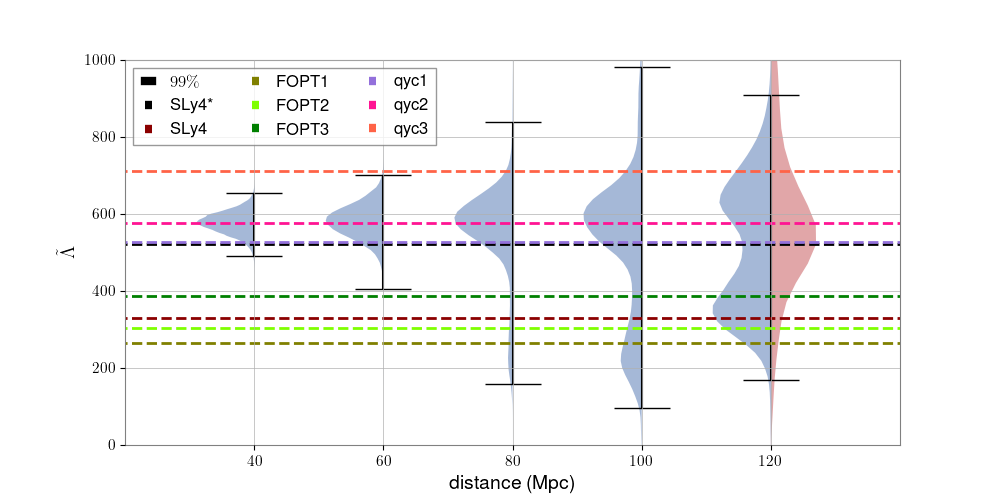}
\caption{In blue, the violin plots represent theprofile of the $\tilde{\Lambda}$ PDF for simulated data with a noise realisation using the O4 PSD at different distances. The injected tidal deformability is calculated from SLy4 EoS (top) and from qyc2 EoS (bottom). In red, the violin plots represent the same thing, but using the average noise over 100 different realisations. The error bars are the 99\% confidence level region. The effective tidal deformabilities corresponding to the EoSs considered are shown by the horizontal dashed lines.} 
\label{fig:violon}
\end{figure}

\subsection{Impact of the distance of the source on O4 signal}
\label{sec:distance}

In Fig.~\ref{fig:violon} we show the impact of the distance on the shape of the reconstructed signal assuming the O4 PSD and SLy4 (top panel) or qyc2 (bottom panel) EoS. The half violin plot in blue represent the PDF of the effective tidal deformability for a single noise realisation. The horizontal dashed lines correspond to the value of the expected effective tidal deformability when a specific EoS is considered (see vertical bars in Fig.~\ref{fig:gw170817_lambda} and in Fig.~\ref{fig:effectivelambda}). The half violin plot in red shows the average PDF obtained from the same simulated event, when superposed to 100 different noise realisations. The red profiles at 40~Mpc and at 120~Mpc with the Sly4 EoS are simply another representation of Fig.~\ref{fig:noise}.  
At a distance of $40$~Mpc, $60$~Mpc, $80$~Mpc, $100$~Mpc and $120$~Mpc, the $99\%$ posterior credible level of $\tilde{\Lambda}$ has respectively a width of $137$, $219$, $517$, $639$ and $740$ when we used SLy4 EoS to generate the simulated data. As a reminder, this interval is equal to $720$ for the $40$~Mpc injection with the O2 PSD.  The posterior profiles get larger with distance, and their 99\% confidence regions, shown in black in the figure, contain all the EoSs from 80~Mpc onwards and do not allow anymore to distinguish between FOPT and quarkyonic transitions to quark matter. At such distance a double-peaked structure starts to appear, getting very similar to the GW170817 signal at about 100 Mpc. This can be interpreted by saying that in O4 we expect that a BNS merger occurring at a distance below about 100 Mpc is more constraining than GW170817 for what concerns the EoS. A BNS merger at a distance of 80(100) Mpc is expected to happen once every $1.5^{+4.5}_{-0.9}$ years ($9^{+27}_{-5}$ months). At 120 Mpc, the observation does not give anymore the ability to prefer one family of EoS over another. It is also surprising to observe that, above 80 Mpc, the analysis of the SLy4 injected signal creates a peak at about twice the expected effective tidal deformability, while the qyc2 injection creates a peak at about one-half the expected tidal deformability.

\section{Conclusions}
\label{sec:conclusions}

In this paper, we have studied the possible constraints on the NS EoS coming from the observation of GW signals from BNS mergers during the LVK O4 observing run, which will start in Spring 2023. In particular, we have considered three scenarios in terms of phase transition, resulting in eight typical EoSs, including 2 nucleonic EoSs (SLy4 and SLy4$^*$), three EoSs with a FOPT (FOPT1, FOPT2, FOPT3) and three EoSs with a (quarkyonic) cross-over to a quark core (qyc1, qyc2, qyc3). The FOPT EoSs show a strong reduction of the radius (and then of the effective tidal deformability) while the quarkyonic ones have an opposite behavior with an increase of the radius for masses compatible with GW170817.

Based on simulation, we have studied the ability to constrain extreme matter EoSs in the future LVK O4 observing run. We have investigated how the improvement of detectors' sensitivities will help in the analysis of an event like GW170817, and our main conclusions are the following:
\begin{itemize}
\item So far it has not been possible to extract any information on the structure of a coalescing NS from a GW signal without considering its electromagnetic counter part. If an event similar to GW170817 occurs during O4 (at about 40 Mpc), the advanced LIGO and Virgo detectors alone will be able to extract a tidal deformability within a 90\% confidence level of about 93, approximately seven times better than for GW170817. With such an event, the Bayes factor allows to sort the EoSs that best fit this simulated event and, with good chance, exclude a certain number of them. 
%effective distance ('H1: ', 112.47420029626824) ('L1: ', 133.21549862631056) ('V1: ', 329.0649712877785)
\item The detection of a single-source in a favorable orientation and located even at larger distances, up to about 100 Mpc, will lead in any case to a better measurement of the tidal deformability, and thus to a sharper EoS selection than what has been possible with GW170817. By combining the results, any new BNS detection will improve our current knowledge on the internal structure of a NS.
\item The recurrent presence of a double peak in the $\tilde{\Lambda}$ posteriors seems to be strongly correlated with the geocenter time and, in turn, to the noise level. Indeed, by using the O4 PSD with a distance less than 80 Mpc, the SNR is large enough to reconstruct the effective tidal deformability with a single peak well centered on the injection value. For larger distances a double peak shape may appear, independently of the EoSs considered, and depending on the specific noise realisation. 
\end{itemize}

\section{Acknowledgment}

This work was granted access to the HPC resources of IDRIS under the allocation 2022-A0120413439 made by GENCI. RS acknowledges support from the Nuclear Physics from Multi-Messenger Mergers (NP3M) Focused Research Hub which is funded by the National Science Foundation under Grant Number 21-16686. The authors are grateful to the LABEX Lyon Institute of Origins (ANR-10-LABX-0066) Lyon for its financial support within the Plan France 2030 of the French government operated by the National Research Agency (ANR). This study is part of a project that has received funding from the European Union’s Horizon 2020 research and innovation program under grant agreement STRONG – 2020 - No 824093 (H.H.).

\appendix

\section{Correlation between the geocentric time and the effective tidal deformability}
\label{app:time}

\begin{figure}[t!]
\begin{subfigure}{0.49\textwidth}
    \includegraphics[width=\textwidth]{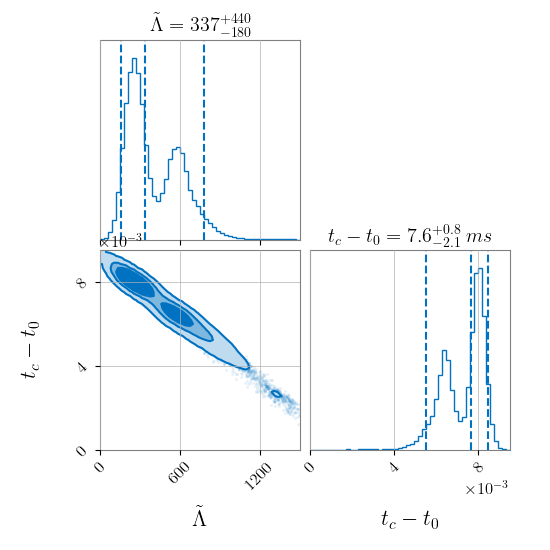}
    \caption{PDF from the {\it flat prior} analysis calculated with GW170817 data.}
    \label{afig:app:tl}
\end{subfigure}
\begin{subfigure}{0.49\textwidth}
\includegraphics[width=\textwidth]{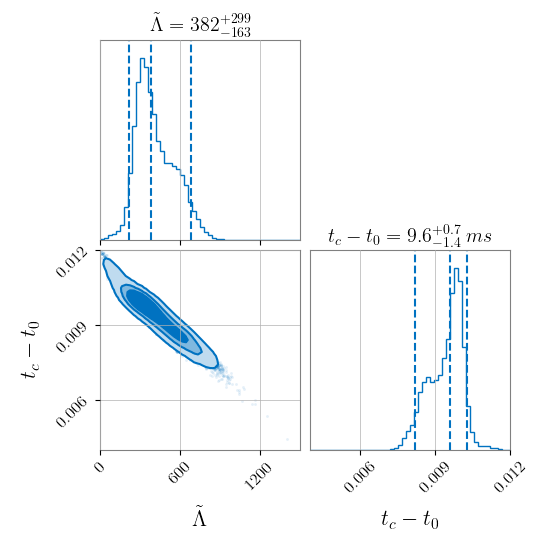}
    \caption{PDF for a simulated signal like GW170817 with O2 sensitivity.}
    \label{bfig:app:tl}
\end{subfigure}
\caption{
%Posterior distributions in the 2-dimensional plane of the geocentric time of merger (with $t_0=1187008882.42$s) versus $\tilde{\Lambda}$ and corresponding 1-dimensional projections. 
Corner plot showing 2 and 1-dimentional marginalised PDF for the geocentric time of merger (with $t_0=1187008882.42$s) and the effective tidal deformability. The 2-dimensional contours show the $68$\%, $90$\% and $99$\% probability regions and the dashed lines on the 1-dimensional plot show the median and the $90$\% probability intervals.}
\label{fig:app:tl}
\end{figure}

We discuss the correlation between the double peak in $\tilde{\Lambda}$ and the double peak distribution in the geocentric time. The noise spectral densities of the current LIGO and Virgo detectors is such that, for BNS signals, the network is mostly sensitive to the inspiral part of the waveform. This affects our capability of constraining the system tidal deformabilities. Assuming different EoSs, corresponding to stiffer or softer NSs, has an effect on the duration of the signal and hence on the determination of the time of the merger. The correlation between tidal deformabilities and time in Fig.~\ref{fig:app:tl} is showing this effect.

\section{Approximation of the Bayes factor}
\label{app:BF}

\begin{figure}[t!]
\centering
\includegraphics[width=0.7\textwidth,trim = {0 52 0 0}, clip]{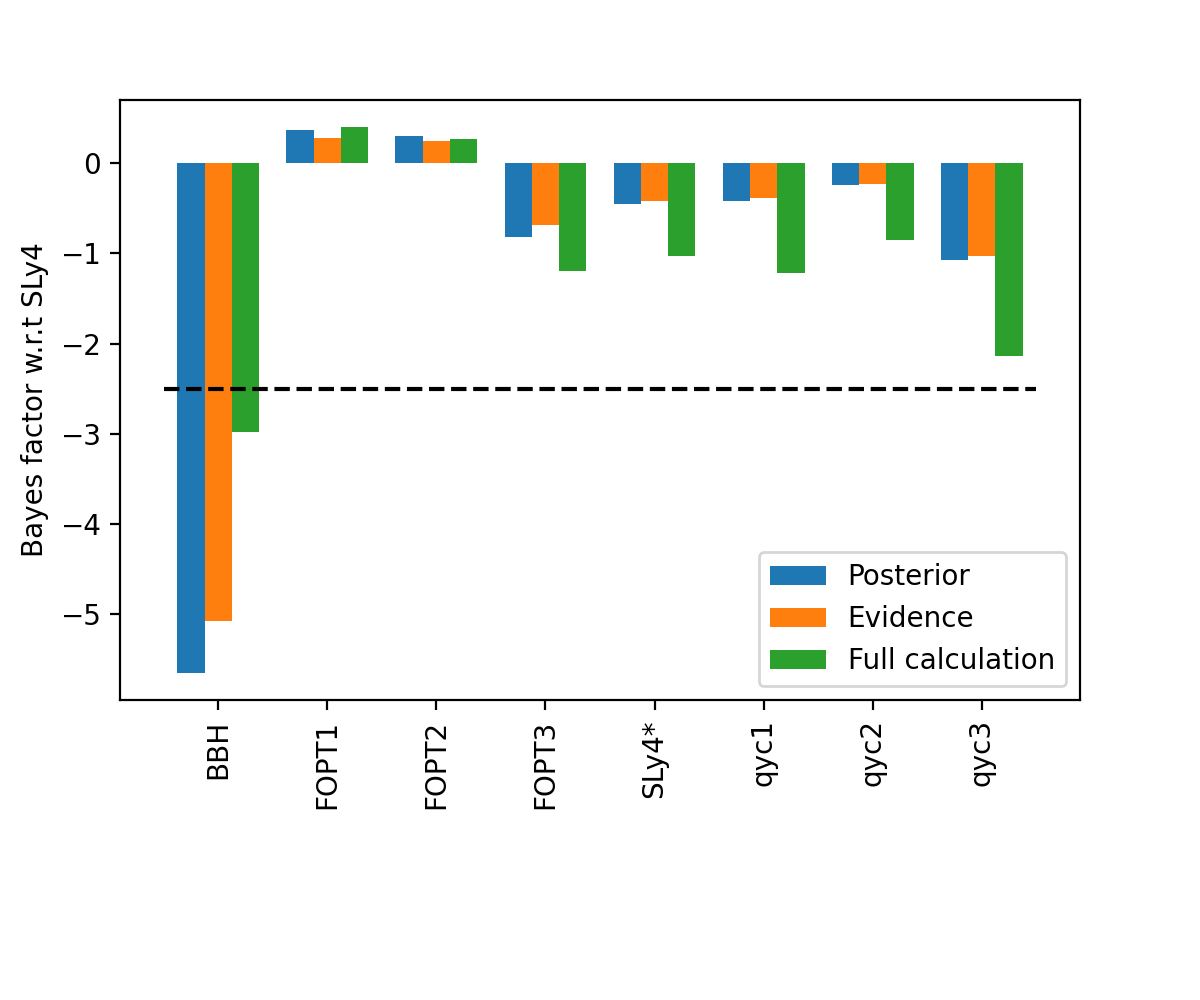}
\caption{Bayes factor for various EoSs considered in this analysis (see horizontal axis for more details) with respect to the SLy4 equation taken as reference. The green histogram represents the results obtained by Bayesian inference of each model. The blue and orange histograms represent respectively two approximate calculations of the Bayes factor from Equations~\eqref{eq:approx_post} and~\eqref{eq:approx_evidence}. The dashed black vertical line is the moderated evidence limit.} 
\label{fig:gw170817_histo}
\end{figure}

For each EoS we considered, three values of the Bayes factor are calculated and shown on the histogram in Fig.~\ref{fig:gw170817_histo}. In green, the exact calculation of the evidence from the Bayesian analysis has been performed. In blue and orange, an approximation of the evidence has been performed allowing an almost instantaneous calculation of the Bayes factor from the simulation named {\it flat prior}. These approximations are detailed in the following. 

The first approximation (so-called Posterior in Fig.~\ref{fig:gw170817_histo}) of the Bayes factor uses the Savage-Dickey density ratio~\cite{stat}. Let us consider two models $A:\mathcal{M}_{\tilde{\Lambda}=\tilde{\Lambda}_0, \theta}$ and $B:\mathcal{M}_{\tilde{\Lambda}, \theta}$ with $\theta$ the set of intrinsic and extrinsic parameters except the tidal deformability which is fixed at $\tilde{\Lambda}_0$ for one of the two models. In a first step, it is assumed that $\delta \tilde{\Lambda}$ intervening at 6pN can be completely neglected compared to $\tilde{\Lambda}$ intervening at 5pN. The Bayes factor can be calculated as~\cite{stat}: 
%\frac{p(\tilde{\Lambda}=\tilde{\Lambda}_0|d,\mathcal{M}_{\tilde{\Lambda}, \theta})\mathcal{Z}(d|\mathcal{M}_{\tilde{\Lambda}, \theta})}{\pi(\tilde{\Lambda}=\tilde{\Lambda}_0|\mathcal{M}_{\tilde{\Lambda}, \theta})} \frac{1}{\mathcal{Z}(d|\mathcal{M}_{\tilde{\Lambda}, \theta})}
\begin{equation}
\mathcal{B}_{AB} = \frac{\mathcal{Z}(d|\mathcal{M}_{\tilde{\Lambda}=\tilde{\Lambda}_0, \theta})}{\mathcal{Z}(d|\mathcal{M}_{\tilde{\Lambda}, \theta})} = \frac{p(\tilde{\Lambda}=\tilde{\Lambda}_0|d,\mathcal{M}_{\tilde{\Lambda}, \theta})}{\pi(\tilde{\Lambda}=\tilde{\Lambda}_0|\mathcal{M}_{\tilde{\Lambda}, \theta})}\,.
\end{equation}
Assuming that the PDF are identical whatever the assumption about nuclear matter and that when an EoS is fixed, $\tilde{\Lambda}$ is perfectly determined $p(\tilde{\Lambda})=\delta(\tilde{\Lambda}-\tilde{\Lambda}_A)$, the Bayes factor can be approximated by the following formula:
\begin{equation}
\mathcal{B}_{AB} = \frac{p(\tilde{\Lambda}=\tilde{\Lambda}_A)}{p(\tilde{\Lambda}=\tilde{\Lambda}_\mathrm{SLy4})}\,.
\label{eq:approx_post}
\end{equation}
$p(\tilde{\Lambda}=\tilde{\Lambda}_A)$ is the $\tilde{\Lambda}$ PDF evaluation from the analysis of {\it flat prior} to the value $\tilde{\Lambda}_A$. This value is the average value of the PDF calculated with the Equation~\eqref{eq:tilde} from the PDF of the chirp mass and mass ratio assuming the tidal deformabilities as a function of NS masses. For each of the models we considered, $\tilde{\Lambda}_A$ is represented by the vertical bars in Figure~\ref{fig:gw170817_lambda}. There is thus a strong correlation between the value of the PDF with a uniform prior evaluated from these vertical bars and the value of the Bayes factor in Figure~\ref{fig:gw170817_histo}. Despite all these approximations, this extremely simple method using only the Bayesian analysis of the {\it flat prior} gives a very good idea of the results. The farthest value from the result is given for the BBH model because the PDF is almost zero at this value. We can add the $\delta \tilde{\Lambda}$ PDF in the formula calculated from the average value of the distribution calculated from $\mathcal{M}$ and $q$ using Equation~\eqref{eq:tilde} but the result does not change much because its PDF is very smooth.    

Another approach (so-called Evidence in Fig.~\ref{fig:gw170817_histo}) to calculate the evidence, assuming the same types of approximation, is given by equation (11) of the article~\cite{Ghosh:2021eqv}: 
\begin{equation}
\mathcal{Z}_A \propto \int p(q', \tilde{\Lambda}_A(\mathcal{M}_0,q')|d)dq'  \,,
\label{eq:approx_evidence}
\end{equation}
with $\tilde{\Lambda}_A$ calculated through the EoSs by the relation shown in figure~\ref{fig:rayon_lambda} and it has been assumed that $\mathcal{M}$ is perfectly determined and equal to $\mathcal{M}_0=1.1975M_{\odot}$. The calculation of the Bayes factor by this method is plotted in orange in the histogram in Figure~\ref{fig:gw170817_histo} and gives very similar values to the method using Equation~\eqref{eq:approx_post}.

These methods, which approximate the Bayes factor for each EoS from a single analysis, give a very good idea of the results but become questionable when the PDF is almost zero. For the set of EoSs considered in this paper, the Bayes factor rank them in the following order: FOPT1, FOPT2, SLy4, qyc2, qyc1, SLy4$^*$, FOPT3, qyc3, BBH from the one best to the worst fitting the data. The same ranking would have been made from the PDF in Figure~\ref{fig:gw170817_lambda} evaluated at the vertical bars.

Let us note however that a new and innovative reduced-order quadrature (ROQ) method is now available allowing faster evaluation of the EoS. Its principle is to build a GW signal based for a set of templates employing PyROQ~\cite{Qi:2020lfr}, which allows a much faster likelihood calculation. This method has already been widely tested for black hole coalescences~\cite{Smith:2016qas} and is also applicable to NS~\cite{Morisaki:2020oqk}. In this case, the Bayesian analysis takes less than a day using only one CPU, instead of 160 CPUs for a day with the usual technique employed in this paper.

\bibliography{biblio}

\end{document}